\documentclass{article}
\usepackage{spconf}

\usepackage{amsmath,amssymb,amsfonts}
\usepackage{graphicx}
\usepackage{textcomp}
\usepackage{xcolor}
\usepackage{signalpreamble}
\def\BibTeX{{\rm B\kern-.05em{\sc i\kern-.025em b}\kern-.08em
    T\kern-.1667em\lower.7ex\hbox{E}\kern-.125emX}}
\begin{document}
\ninept

\title{Optimal Sensor Placement via Graph-Constrained Flow Matching}
\name{Feng Ji, Jingyang Dai, Wee Peng Tay, Sirajudeen Gulam Razul \thanks{F. Ji and J. Dai contribute equally.}}
\address{
\fontsize{10}{10}\selectfont
School of Electrical and Electronic Engineering, Nanyang Technological University
}

\maketitle

\begin{abstract}
Optimal sensor placement is a fundamental problem in graph signal processing (GSP), where a limited number of sensors are deployed to reconstruct a continuous signal field. Existing GSP methods rely on combinatorial optimization over discretized graphs, resulting in high computational cost and sensor locations restricted to graph vertices. We reformulate sensor placement as a continuous-space generative modeling problem. An offline GSP optimization routine first generates training samples of optimal sensor configurations, from which a flow matching (FM) model learns their distribution. At inference, the learned velocity field directly generates continuous sensor coordinates, eliminating online combinatorial optimization. We further develop a permutation-invariant conditional generation framework for deployment with fixed anchor sensors. Experiments on a realistic radio propagation simulator demonstrate the effectiveness of the proposed framework for sensor placement.
\end{abstract}


\section{Introduction} \label{sec:int}
Optimal sensor placement is a foundational problem in network engineering and graph signal processing (GSP), with critical applications ranging from environmental monitoring and infrastructure security to wireless communications \cite{Shu13, Sah22, Ma25}. Given a bounded continuous domain, we consider the task of identifying a sparse subset of spatial locations to deploy a (small) fixed budget of sensors such that the global, continuous signal field can be reconstructed with minimal error. Traditional approaches typically discretize the spatial domain into an ambient graph and invoke greedy combinatorial optimization algorithms \cite{Ani16, Cha18}, such as sequential least squares or frame potential maximization, to identify highly informative nodes. 

Moreover, in practical deployment scenarios, sensor network design is rarely an unconditional, greenfield task. Instead, engineers are frequently confronted with a \emph{conditional deployment problem}, which manifests across several modern applications:
\begin{itemize}
    \item Adaptive network expansion \cite{Ber23}: An operational region already possesses a set of $k_1$ legacy, static sensors, and an operator seeks to deploy $k - k_1$ additional sensors to maximize the expanded network's reconstruction fidelity under changing spatial field dynamics.
    \item Autonomous drone swarm coordination \cite{Cha25}: In mobile sensing or search-and-rescue operations using UAV or drone swarms, a subset of $k_1$ sensing drones may have their positions fixed due to local topographical tracking constraints, battery emergencies, or waypoint hovering. The system must instantaneously generate the coordinates for the remaining $k - k_1$ drones to ensure optimal spatial coverage of the signal field.
    \item Obstacle-aware dynamic relocation \cite{Ale24}: In complex geometric environments, localized interference might render certain regions blind, requiring adaptive positioning of mobile nodes relative to fixed, unmovable anchor basestations.
\end{itemize}

Solving these placement problems using classical GSP has a few limitations. It requires re-running combinatorial optimization over the entire graph for every unique anchor configuration, and the complexity often scales as a function of the graph size $n$ rather than merely the sensor number $k \ll n$. In privacy- or security-sensitive applications, this repeated optimization may require repeated access to the underlying signal data, increasing the risk of unintended information disclosure. Furthermore, discrete algorithms cannot place sensors in the continuous spatial intervals between predefined graph vertices, potentially limiting the network’s performance.

To bypass these limitations, we propose a shift in paradigm: rather than treating sensor placement as an online discrete optimization problem, we frame it as a continuous-space generative inference problem. We introduce a framework that leverages \emph{flow matching (FM)} \cite{lip23} over continuous coordinate point clouds. By utilizing an accelerated, regularized GSP oracle strictly during the offline training phase, our generative model learns the underlying continuous probability manifold of optimal sensor configurations. At inference time, the learned deterministic velocity fields guide a set of random noise particles toward optimal sensing coordinates, and the process does not depend on the graph (of size $n$) anymore.

To the best of our knowledge, this is the first work that bridges GSP and continuous generative modeling based on FM for this type of task. Our core contributions are as follows:
\begin{itemize}
    \item We formulate a \emph{continuous-space generative framework} for sensor placement via \emph{GSP-generated supervision} that breaks the grid-dependency of traditional graph-based optimization, allowing for super-resolution, off-grid inference.
    \item We introduce a \emph{permutation-invariant conditional inference} pipeline, ensuring that the model seamlessly accommodates pre-fixed physical anchors without retraining.
\end{itemize}
The efficacy of our method is validated via a realistic radio propagation simulator.

\section{Preliminaries}
\subsection{Graph signal processing} \label{sec:wbr}
In this subsection, we briefly review GSP \cite{Shu13, Ort18, JiaFenTay:J23} concepts most relevant to this paper, with emphasis on signal sampling and recovery.

Consider an undirected weighted graph $G=(V,E)$ of size $n=|V|$. Let $\bL$ denote its weighted graph Laplacian, which admits the eigendecomposition $\bL=\bU\bLambda\bU^\top$, where the columns of $\bU$ form an orthonormal eigenbasis of $\bL$. A graph signal $\bx\in\mathbb{R}^n$ assigns a scalar value to each node in $V$. It is said to be $w$-bandlimited if all but $w$ coefficients of its graph Fourier transform $\widehat{\bx} = \bU^\top\bx$ are zero. Under this assumption, $\bx$ can be perfectly reconstructed from its values observed at $w$ appropriately selected nodes.

Specifically, let $V_0\subset V$ be a set of $w$ sampled nodes. Denote by $\bx_{V_0}$ the subvector of $\bx$ corresponding to the nodes in $V_0$, and by $\bU_{V_0}$ the $w\times n$ submatrix of $\bU$ whose rows correspond to the nodes in $V_0$. If $\bx$ is $w$-bandlimited and $\bU_{V_0}^\top\bU_{V_0}$ is invertible, then $\bx$ can be recovered exactly from $\bx_{V_0}$ via
\begin{align} \label{eq:xuu}
\bx = \bU(\bU_{V_0}^\top\bU_{V_0})^{-1}\bU_{V_0}^\top\bx_{V_0}.
\end{align}
Motivated by the deployment of a limited number of sensors for signal reconstruction over a large spatial region, we use the above formula as the GSP input for our proposed model.

\subsection{Flow matching}\label{sec:fm}
Flow matching (FM) \cite{lip23} is a generative modeling framework designed to transform a simple source distribution (e.g., standard Gaussian noise) into a complex target data distribution. The core intuition behind FM is to find a time-dependent vector field whose paths seamlessly join the source and target distributions. By following this vector field over a time interval $t \in [0, 1]$, noise points are smoothly pushed along \emph{trajectories $\phi_t$} until they arrive exactly at the data distribution.

Formally, let $x \in \mathbb{R}^d$ be the state variable. One defines a vector field $u: [0, 1] \times \mathbb{R}^d \to \mathbb{R}^d$ that acts as a velocity field. A starting point $x \sim \mu_0$ from the noise distribution moves over time along $\phi_t$ according to the ordinary differential equation:
\begin{align} \label{eq:ddt}
    \frac{d}{dt} \phi_t(x) = u_t(\phi_t(x)), \quad \phi_0(x) = x.
\end{align}
At $t=1$, the final point $\phi_1(x)$ represents a sample from the generated data distribution $\mu_1(x)$.

To train a neural network $u_\theta(t, x)$ to approximate this ideal vector field, one minimizes the difference between our model and a target vector field $u_t(x)$:
\begin{align} \label{eq:lft}
    \mathcal{L}_{\text{FM}}(\theta) = \mathbb{E}_{t, \mu_t(x)}  \| u_\theta(t, x) - u_t(x) \|^2.
\end{align}
The global target vector field $u_t(x)$ depends on the entire data distribution and is intractable. To resolve this, one conditions the trajectories on final data points $x_1 \sim \mu_1(x)$. This allows one to define straight paths from the source $x$ to $x_1$: $x_t = t x_1 + (1-t)x$.

Since the path is a constant speed straight line, the ideal conditional velocity is independent of time and trivial to calculate: $u_t(x \mid x_1) = x_1 - x$. The training objective simplifies to interpolating between noise and data, and training the neural network to predict this constant direction vector $x_1 - x$ at any intermediate time $t$.

\subsection{Conditional sensor placement}
We now formally define the problem to be addressed. Consider a continuous region\footnote{Our approach also applies to $\Omega$ being in a general Riemannian manifold.} $\Omega \subset \mathbb{R}^2$ with an underlying signal distribution $\mu$ over the \emph{signal space} $S(\Omega)$, where each signal is a function on $\Omega$. For $\boldf \sim \mu$ and a set of $k$ sensor locations $\mathcal{P} = \{p_1, \dots, p_k\} \subset \Omega$, we observe the discrete signal $\boldf_{\calP} = (\boldf(p_i))_{i\leq k} \in \mathbb{R}^k$. A problem of interest in signal processing is to find a signal reconstruction function $\Psi$ from these observations such that
\begin{align} \label{eq:ebs}
\mathbb{E}_{\boldf\sim \mu}\calL\big(\Psi(\boldf_{\calP}), \boldf\big)
\end{align}
is small for a prescribed loss function $\calL$. In this paper, we consider the sensor placement problem, which further seeks the sensor locations $\calP$ that minimize the expected reconstruction error in \cref{eq:ebs} for given $\Psi$ and $\calL$.

As discussed in \cref{sec:int}, many practical applications require the more general setting of conditional deployment. Specifically, suppose that a subset of $k_1$ sensors is fixed at anchor locations $\mathcal{P}_{\mathrm{fix}} \subset \Omega$. The goal is to determine the remaining $k-k_1$ sensor locations such that the same optimization objective is achieved.

To briefly preview our strategy, instead of solving \cref{eq:ebs} directly, which is generally intractable for arbitrary $\mu$, we adopt a flexible, generative approach. It learns a distribution over sensor locations via solving \cref{eq:ebs} with $\mathbb{E}_{\boldf\sim\mu}$ replaced by an empirical average. The distribution assigns a higher probability to configurations with superior reconstruction performance. Sensor deployments are then generated by sampling from this learned distribution.

To this end, we first employ GSP to construct a training set for unconditional sensor placement using signals sampled from $\mu$, where the signal reconstruction function $\Psi$ is derived according to \cref{sec:wbr}. We then train an FM model on these optimized sensor configurations to generate high-quality sensor placements for both the unconditional and anchor-constrained settings.

\section{Methodology}
\subsection{Empirical sampling via graph signal processing} \label{sec:esv}
\begin{figure}[!htb]
    \centering
    \includegraphics[width=1\linewidth]{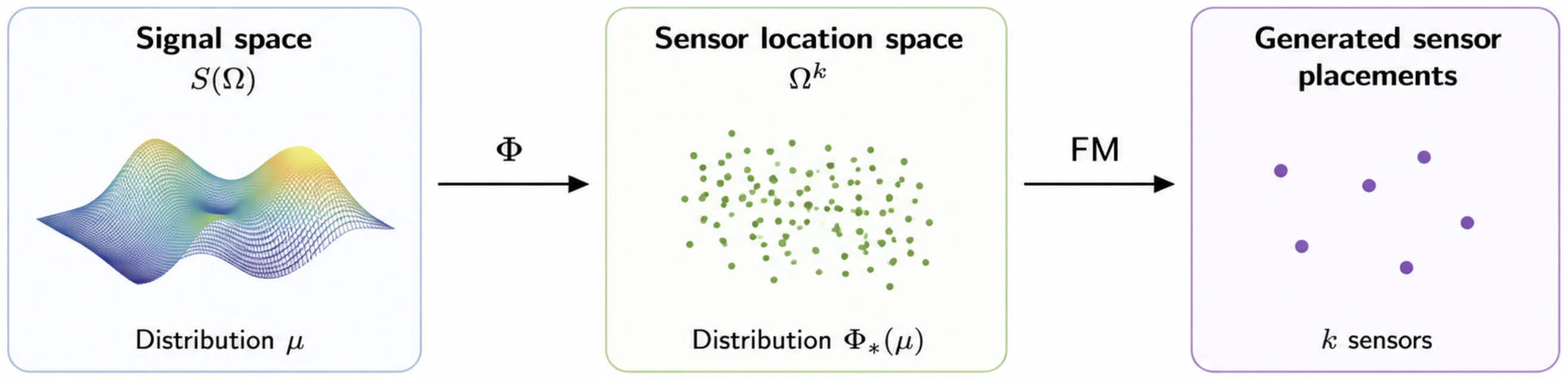}
    \caption{$\Phi$ pushes the signal distribution to a distribution of sensor locations, from which we obtain samples to train the FM model. }
    \label{fig:ioo}
\end{figure}

We first outline the overall design of the proposed framework (see \cref{fig:ioo}). Specifically, we construct a (measurable) function on the signal space $\Phi: S(\Omega) \to \Omega^k$, where the codomain $\Omega^k$ represents the locations of $k$ sensors. Intuitively, $\Phi$ maps an input signal $\boldf$ to the sensor locations $\calP$ whose observations $\boldf_\calP$ yield the best reconstruction of $\boldf$ under a GSP-based reconstruction scheme. The function $\Phi$ then pushes forward the signal distribution $\mu$ on $S(\Omega)$ to a distribution $\Phi_*(\mu)$ on $\Omega^k$, i.e., for any measurable subset $U \subset \Omega^k$, $\Phi_*(\mu)(U) = \mu(\Phi^{-1}(U))$. Consequently, given samples from $\mu$, the function $\Phi$ generates corresponding samples of sensor locations in $\Omega^k$, which are subsequently used to train an FM model for general-purpose sensor deployment, including anchor-constrained settings.   

As the domain $\Omega$ is continuous, we first discretize it to incorporate GSP. Specifically, we construct an ambient graph $G=(V,E)$, whose size satisfies $n\gg k$, via a \emph{Voronoi tessellation} \cite{Aur91} that closely approximates $\Omega$. Let $V \subset \Omega$ be a set of $n$ points forming an \emph{$\epsilon$-net}, i.e., for every $x \in \Omega$, there exists $v\in V$ such that $d(x,v) \leq \epsilon$. A sufficiently small hyperparameter $\epsilon$ ensures that $V$ provides a faithful approximation of $\Omega$.

Given an $\epsilon$-net $V=(v_1,\ldots,v_n)$, we construct the Voronoi tessellation of $\Omega$ as follows. For each $v_i\in V$, the (convex) \emph{cell} $C_i$ centered at $v_i$ consists of all $x\in \Omega$ satisfying $d(x,v_i) \leq d(x,v_j)$ for every $i\neq j\leq n$. The cells form a partition of $\Omega$ and intersect only along their boundaries. We then construct the graph $G=(V,E)$ with node set $V$, where two nodes $v_i,v_j\in V$ are connected by an edge whenever $C_i\cap C_j\neq \emptyset$. Edge weights are assigned using the Gaussian kernel $w_{i,j} = \exp(-d(v_i,v_j)^2/2c^2)$ for some parameter $c$. The corresponding graph Laplacian is denoted by $\bL = \bU\bLambda\bU^\top$, following the notation in \cref{sec:wbr}.

Each signal $\boldf \in S(\Omega)$ naturally induces a graph signal $\bx$ on $G$ by setting $\bx(i) = \boldf(v_i)$. For simplicity, we also denote the induced graph signal distribution by $\mu$.

Having established the graph representation, we now apply the GSP framework reviewed in \cref{sec:wbr}. Given $m$ graph signals $\bx_1,\ldots,\bx_m\sim \mu$, we employ a \emph{greedy sampling algorithm} \cite{Cha18} to construct a subset $\calP \subset V$ of sensor locations. We initialize $\calP_0=\emptyset$. At the $\ell$-th iteration, where $\ell\leq k$, we update $\calP_\ell = \calP_{\ell-1} \cup \{v^*\}$, where $v^*$ is obtained by solving
\begin{align} \label{eq:vvv}
v^* = \argmin_{v\in V\backslash \calP_{\ell}} \sum_{j=1}^m \norm{\bx_j - \bU\big(\bU_{V_{\ell}}^\top\bU_{V_{\ell}}+\lambda \bI\big)^{-1}\bU_{V_{\ell}}^\top\bx_{j,V_{\ell}}}^2,
\end{align}
here we write $V_{\ell}=\calP_{\ell-1}\cup \{v\}$ for convenience. After $k$ iterations, the selected sensor locations are given by $\calP$.

\emph{Discussions}. We note that \cref{eq:vvv} is obtained by modifying \cref{eq:xuu} with the regularization term $\lambda\bI$ to ensure matrix invertibility. Consequently, the procedure directly applies the GSP signal reconstruction framework described in \cref{sec:wbr}, using the corresponding signal reconstruction function $\Psi$ and reconstruction loss $\calL$ in \cref{eq:ebs}. Thus, the greedy algorithm addresses \cref{eq:ebs} when $\mathbb{E}_{\boldf\sim\mu}$ is replaced with the average over samples $\bx_1,\ldots,\bx_m$. Furthermore, the computation can be accelerated by updating the matrix inverse via the Sherman–Morrison formula \cite{She50}.

The above procedure represents the function 
\begin{align*}
\Phi: S(\Omega) \to V^k \subset \Omega^k
\end{align*}
outline at the beginning of this subsection. It is repeated for $T$ iterations, producing a collection of sensor configurations $\calS = \{\calP^{(1)},\ldots,\calP^{(T)}\}$. These samples are from the pushforward distribution $\Phi_*(\mu)$ (as the target distribution) and constitute the training set for the FM model described in \cref{sec:fm}. In the next subsection, we discuss a few outstanding technical issues when applying FM. 

\subsection{Sensor placement via flow matching} \label{sec:spv}
Recall that we have obtained a collection of sensor configurations $\calS = \{\calP^{(1)},\ldots,\calP^{(T)}\}$. Replacing the graph nodes in each $\calP^{(i)}$ with their corresponding spatial coordinates, we regard the target distribution of the generative model as being supported on $\Omega^k$, or $\Omega^{k-k_1}$ when $k_1$ anchor sensors are fixed. Although FM has been reviewed in \cref{sec:fm}, we provide additional technical details here to address three practical issues: (a) permutation invariance of sensor configurations; (b) anchor-constrained sensor placement; and (c) the encoder architecture for the vector field $u_{\theta}$ in \cref{eq:lft}.

Let $\calP_{\text{fix}} = \{p_1^*,\ldots,p_{k_1}^*\}$ denote the fixed anchor sensor locations. Starting from a noisy sample $\calP_0=\{p_{0,1},\ldots,p_{0,k}\}$, our goal is to generate $\calP_1=\{p_{1,1},\ldots,p_{1,k}\}$ such that $\calP_{\text{fix}}$ is a subset of $\calP_1$. Since sensor configurations are unordered sets whereas FM generates an ordered sequence of coordinates, it is unclear which generated sensors should correspond to the prescribed anchors.

To resolve this ambiguity, we perform a spatial matching step at $t=0$. Specifically, we solve the \emph{linear assignment problem} (e.g., using the Hungarian algorithm \cite{Kuh55}):
\begin{align*}
\pi^* = \argmin_{\pi} \sum_{j=1}^{k_1}  \norm{p_{0,\pi(j)} - p_j^*}^2,
\end{align*}
where $\pi$ is a permutation of the $k$ sensor indices. This assigns each anchor to its nearest sensor in the initial noisy configuration, allowing the ODE trajectories in \cref{eq:ddt} to evolve along paths of minimal displacement while avoiding undesirable global index-shuffling artifacts.

Having established the correspondence, we henceforth treat the sensor sets $\calP_t$, $t\in[0,1]$, as ordered. Following the assignment $\pi^*$, the matched sensors are gradually guided toward their target anchors through a \emph{sigmoidal blending schedule} \cite{Rod19}:
\begin{align*}
\sigma_t = \big(1 + e^{-\kappa(t - t_{0})}\big)^{-1},
\end{align*}
which interpolates between the FM-generated sensor location and the target anchor location. The modified trajectory for a matched sensor $i=\pi^*(j) \in \text{Im}(\pi^*)$ at time $t$ is
\begin{align*}
p_{t,i} = (1 - \sigma_t) \hat{p}_{t,i} + \sigma_t p_{j}^*,
\end{align*}
where $\hat{p}_{t,i}$ denotes the sensor location generated by FM at time $t$. Choosing $t_0=0.5$ and $\kappa=10$ ensures that the sensors evolve almost freely at the beginning of the trajectory, while satisfying $p_{1,i}\approx p_j^*$ at $t=1$ as required. The remaining $k-k_1$ sensors evolve freely throughout the entire process, and together with the anchor sensors constitute the final sensor configuration $\calP_1$.

While we have completed the overall framework, it remains to explicitly describe the encoder architecture for the velocity field $u_{\theta}$ in \cref{eq:lft}. We model the velocity field using a \emph{point transformer} \cite{Vas17}. Specifically, each sensor coordinate $p_{t,i}$ interacts with the entire sensor configuration through a multi-head attention:
\begin{align*}
    \bh_{t,i} = \text{Attention}\Big(&\text{Query}(p_{t,i}, t), \text{Keys}(\{p_{t,j}\}_{j\leq k}, t), \\
    &\text{Values}(\{p_{t,j}\}_{j\leq k}, t)\Big).
\end{align*}
The resulting feature vector is then passed through an \text{MLP}, which outputs the velocity vector $u_{\theta}(t,p_{i,t})$ at $p_{i,t}$. The parameter set $\theta$ comprises all parameters in the multi-head attention module together with those of the final \text{MLP}.

By omitting positional encodings, the model is permutation equivariant, ensuring that the velocity of each sensor depends solely on its spatial relationship with the remaining sensors in the configuration.

\section{Numerical results}
\begin{figure}[!htb]
    \centering
    \includegraphics[width=1\linewidth]{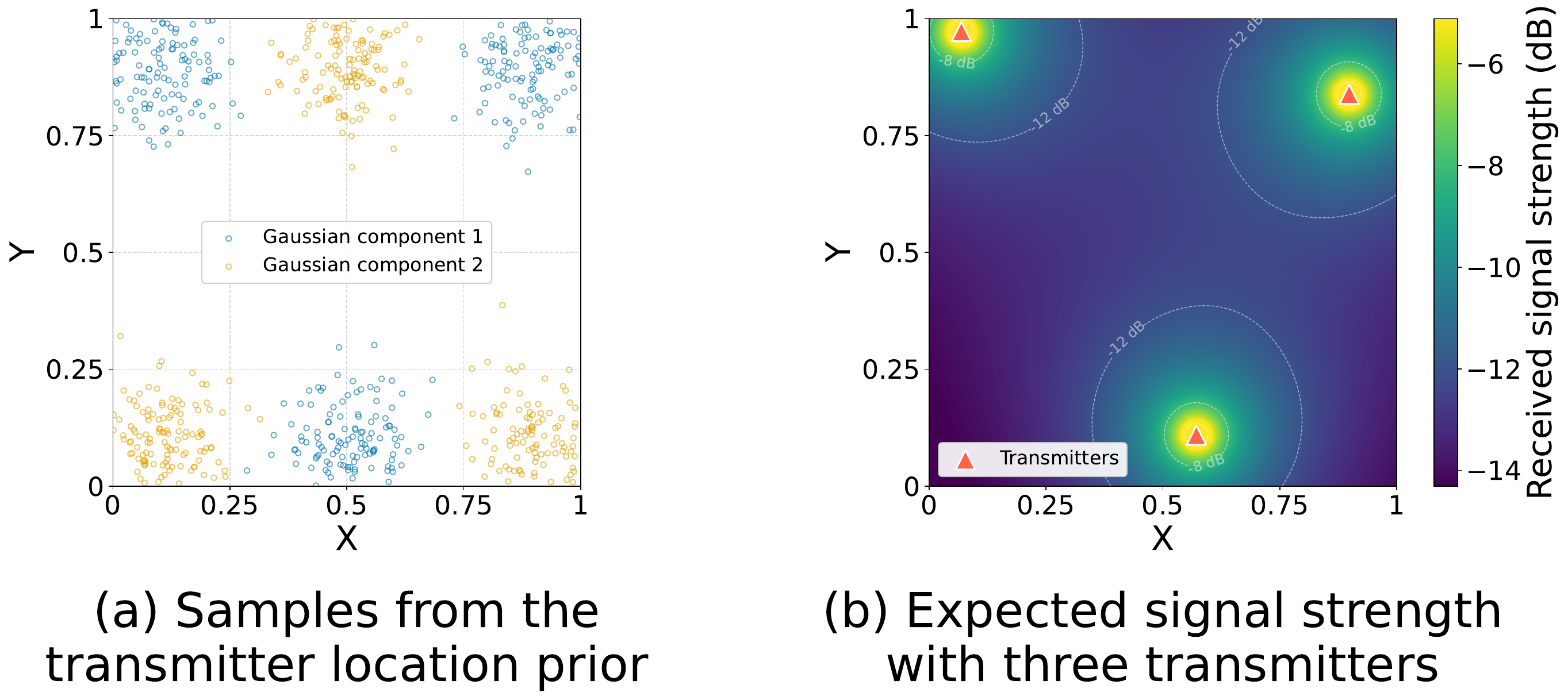}
    \caption{In (a), we show the empirical approximation of a two-peak Gaussian mixture for the locations of $3$ transmitters, called ``triangle''. In (b), we give an example of the sample locations of the transmitters (from the Gaussian mixture) and the resulting signal over $\Omega$.}
    \label{fig:itl}
\end{figure}

In the numerical study, we consider the \emph{communication sensor network} described in \cite{Jia25}. Specifically, there is a flat 2D area $\Omega$ with 3 wireless transmitters, and their locations jointly follow a mixed Gaussian distribution with two peaks (see \cref{fig:itl}(a) for an illustration). The signal received at each location from a single transmitter is subject to path loss, shadow fading, and fast fading. The explicit formula can be found in \cite{Jia25, Cai03}, and we show an example in \cref{fig:itl}(b). Furthermore, the received signal is subject to additive Gaussian noise. The uncertainty in the transmitter locations and signal reading collectively gives the signal distribution $\mu$. 

\begin{figure}
    \centering
    \includegraphics[width=0.45\linewidth]{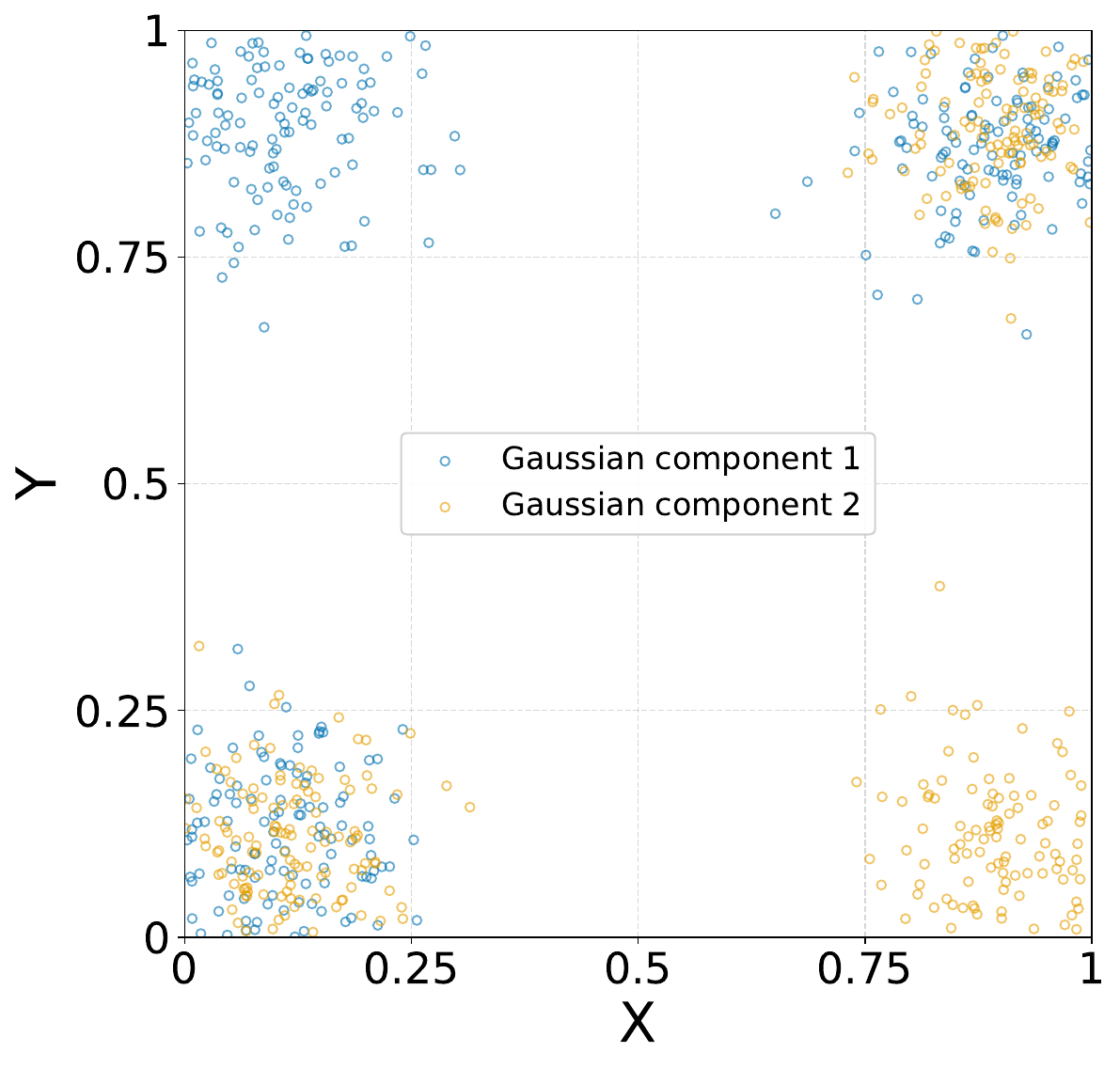}
    \includegraphics[width=0.45\linewidth]{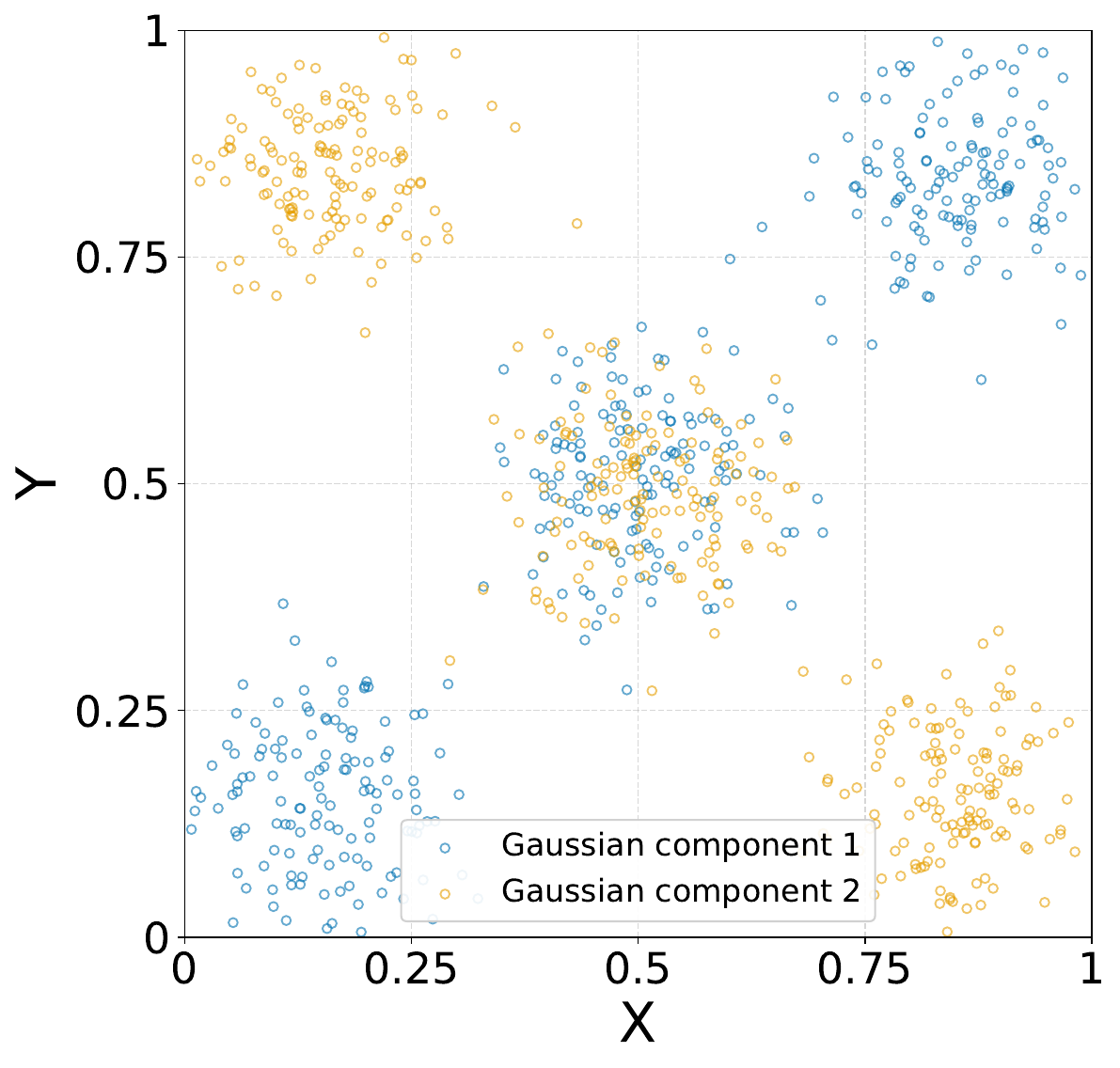}
    \caption{We show the transmitter joint distributions ``Lshape'' and ``diagonal'' used for evaluation.}
    \label{fig:wst}
\end{figure}

In this study, we want to place $10$ sensors to measure and reconstruct the communication signals, and further assume $5$ of them are pre-determined anchors whose locations are generated according to different schemes as discussed below.  

To implement our scheme, we let $V$ be the uniformly spaced $100$ points in $\Omega$, and the Voronoi tessellation generates the standard $10\times 10$ lattice graph $G=(V,E)$. Using GSP outlined in \cref{sec:esv}, we generate a collection of sensor configurations $\calS$ with $T = 1600$ samples. FM described in \cref{sec:spv} allows us to identify the locations of $5$ sensors to be optimally placed. In addition, we generate another collection $\calS_0$ of $200$ configurations for testing and evaluation. 

We compare the following experimental settings:
\begin{itemize}
    \item GB: The \underline{g}olden \underline{b}aseline $S_0$.  
    \item SF: The anchor sensor locations are chosen from \underline{s}amples in $\calS_0$, and the remaining sensors are deployed according to \underline{F}M in \cref{sec:spv}. Under this setting, the anchors are reasonably placed according to the signal distribution. 
    \item SP: The anchor sensor locations are chosen from \underline{s}amples in $\calS_0$, and the remaining sensors are deployed using \underline{P}RODIGY \cite{Sha24}, a benchmark generative model under constraints. 
    \item RF: The anchor sensor locations are \underline{r}andom, and the remaining sensors are deployed according to \underline{F}M in \cref{sec:spv}. Under this setting, the anchors can be suboptimally placed as there is no input from the signal distribution. 
    \item SR: The anchor sensor locations are chosen from \underline{s}amples in $\calS_0$, and the remaining sensors are deployed \underline{r}andomly.
    \item AR: \underline{A}ll sensors are placed \underline{r}andomly. 
\end{itemize}

Notice that the generated locations have continuous coordinates. For evaluation, we project them to their nearest node in the graph $G$, and the resulting node set is $V_{10}$. We consider the following two metrics: 
\begin{itemize}
    \item \emph{A-optimality} (cf.\ \cref{eq:vvv}): $\text{tr}\big((\bU_{V_{10}}^\top\bU_{V_{10}}+\lambda\bI)^{-1} \big)$. This is a generic metric for bandlimited graph signals \cite{Ani16}. It does not take input from the distribution, but relies on the observation that the resulting signals on $G$ are usually smooth, as shown in \cref{fig:itl}(b).    
    \item Empirical \emph{relative mean squared error} (RMSE) for reconstruction: we draw signals $\bx_1,\ldots,\bx_{50}$ according to $\mu$. Let $\calP^*$ be the sensor configuration solved according to the greedy algorithm in \cref{sec:esv} without constraints, treated as the ground truth. Its squared reconstruction error (the summation in \cref{eq:vvv} for $m=50$) is denoted by $\epsilon^*$. For each of the above settings, we compute the corresponding squared reconstruction error as $\epsilon$. Then the RMSE is $\epsilon/\epsilon^*$.    \end{itemize}   
\begin{table}[!htb]
\centering
\caption{A-optimality gap: difference with GB; and RMSE. The best result other than GB in each row is {\bf boldfaced}.} \label{tab:gsr}
\centering
\begin{tabular}{ccccccc} 
\toprule
&GB &SF &SP &RF &SR &AR \\ 
\midrule
triangle& 0& {\bf 13.2} & 20.6& 38.4& 119.6& 196.3 \\ 
Lshape& 0& {\bf 11.8} & 12.6& 36.7& 115.8& 193.3\\ 
diagonal& 0& {\bf 17.9} & 22.7& 43.9& 149.7& 192.8 \\ 
\midrule
triangle& 1.15& {\bf 1.27}& 1.46& 1.46 & 1.97& 2.25 \\ 
Lshape& 1.15& {\bf 1.26}& 1.32& 1.47& 2.57& 1.98\\ 
diagonal& 1.23& 1.35& {\bf 1.34}& 1.93& 2.54& 2.59 \\ 
\bottomrule
\end{tabular}
\end{table}

\begin{figure}[!htb]
    \centering
    \includegraphics[width=0.9\linewidth]{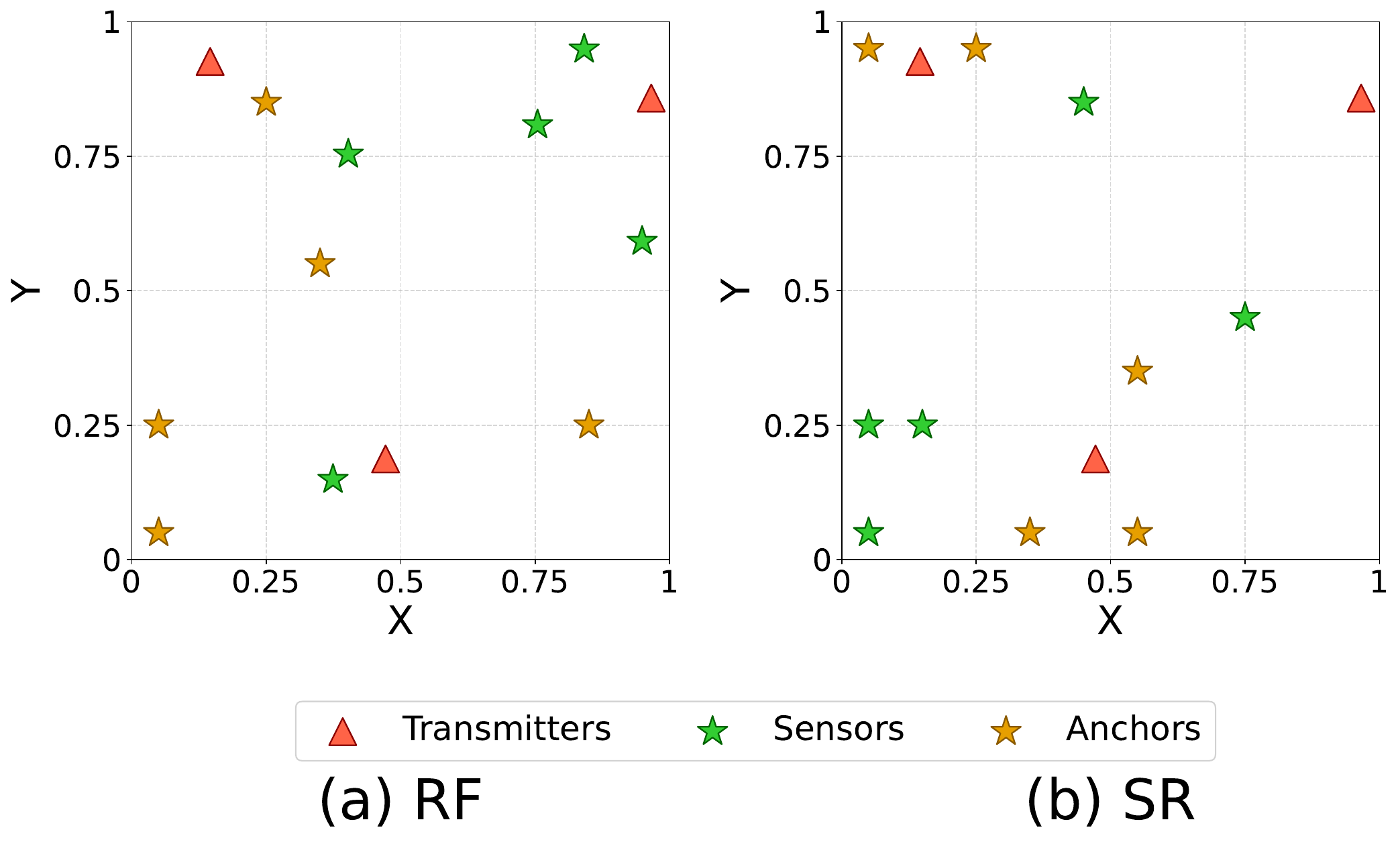}
    \caption{In this example, RF places sensors to effectively ``fill in the gaps’’ of the anchor configuration (e.g., around the top-right transmitter). In contrast, SR cannot consistently achieve this, as the additional sensors are placed randomly.}
    \label{fig:rps}
\end{figure}

We run the experiments for $3$ distinct distributions of the transmitter locations in \cref{fig:itl,fig:wst}: triangle, Lshape, and diagonal. The results, averaged over $200$ runs, are summarized in \cref{tab:gsr}. For either metric, we see that in general, being close to GB, our proposed model performs the best when the anchors are from $\calS_0$. This suggests that our generative approach almost faithfully recovers the sensor location distribution, thereby selecting the best possible locations. It is also interesting to notice that RF always outperforms SR, though RF has suboptimal anchor locations. This suggests that our model rectifies the suboptimality of the constraints via the dynamics of the generative process for the newly deployed sensors (see \cref{fig:rps} for an example). In summary, the learned distribution compensates for poor anchors.   

\section{Conclusions}
We proposed a continuous-space generative framework for sensor placement that leverages GSP-generated training samples and flow matching to eliminate online combinatorial optimization. The proposed method supports both unconditional and anchor-constrained deployment through a permutation-invariant conditional generation framework. Experimental results on realistic radio propagation data demonstrate its effectiveness and flexibility for continuous sensor placement.

\bibliographystyle{IEEEtran}
\bibliography{bib/ref}

@Article{Shu13,
author   = {D. I. Shuman and S. K. Narang and P. Frossard and A. Ortega and P. Vandergheynst},
title    = {The emerging field of signal processing on graphs: Extending high-dimensional data analysis to networks and other irregular domains},
journal  = {IEEE Signal Process. Mag.},
year     = {2013},
volume   = {30},
number   = {3},
pages    = {83-98},
}

@ARTICLE{Cha18,
  author={Chamon, Luiz F. O. and Ribeiro, Alejandro},
  journal={IEEE Trans. Signal Process.}, 
  title={Greedy Sampling of Graph Signals}, 
  year={2018},
  volume={66},
  number={1},
  pages={34-47},
 }

@article{She50,
 author = {Jack Sherman and Winifred J. Morrison},
 journal = {Ann. Math. Stat.},
 number = {1},
 pages = {124--127},
 title = {Adjustment of an Inverse Matrix Corresponding to a Change in One Element of a Given Matrix},
 volume = {21},
 year = {1950}
}

@article{Kuh55,
author = {Kuhn, H. W.},
title = {The Hungarian method for the assignment problem},
journal = {Nav. Res. Logist. Q.},
volume = {2},
number = {1-2},
pages = {83-97},
year = {1955}
}

@Article{Rod19,
AUTHOR = {Sergio Rodrigues, Paulo and Wachs-Lopes, Guilherme and Morello Santos, Ricardo and Coltri, Eduardo and Antonio Giraldi, Gilson},
TITLE = {A q-Extension of Sigmoid Functions and the Application for Enhancement of Ultrasound Images},
JOURNAL = {Entropy},
VOLUME = {21},
YEAR = {2019},
NUMBER = {4},
}

@inproceedings{lip23,
      title={Flow Matching for Generative Modeling}, 
      author={Yaron Lipman and Ricky T. Q. Chen and Heli Ben-Hamu and Maximilian Nickel and Matt Le},
      booktitle = {ICLR},
      year={2023},
}

@inproceedings{Vas17,
  author       = {Ashish Vaswani and
                  Noam Shazeer and
                  Niki Parmar and
                  Jakob Uszkoreit and
                  Llion Jones and
                  Aidan N. Gomez and
                  Lukasz Kaiser and
                  Illia Polosukhin},
  title        = {Attention Is All You Need},
booktitle = {NeurIPS}, 
  year         = {2017},
}

@ARTICLE{Jia25,
  author={Jian, Xingchao and Gölz, Martin and Ji, Feng and Tay, Wee Peng and Zoubir, Abdelhak M.},
  journal={IEEE Trans. Signal Process.}, 
  title={A Graph Signal Processing Perspective of Network Multiple Hypothesis Testing With False Discovery Rate Control}, 
  year={2025},
  volume={73},
  number={},
  pages={3496-3512}
  }

@ARTICLE{Cai03,
  author={Xiaodong Cai and Giannakis, G.B.},
  journal={IEEE Trans. Veh.
Technol}, 
  title={A two-dimensional channel simulation model for shadowing processes}, 
  year={2003},
  volume={52},
  number={6},
  pages={1558-1567},
}

@ARTICLE{Ani16,
  author={Anis, Aamir and Gadde, Akshay and Ortega, Antonio},
  journal={IEEE Trans. Signal Process.}, 
  title={Efficient Sampling Set Selection for Bandlimited Graph Signals Using Graph Spectral Proxies}, 
  year={2016},
  volume={64},
  number={14},
  pages={3775-3789},
}

@inproceedings{Sha24,
author = {Sharma, Kartik and Kumar, Srijan and Trivedi, Rakshit S},
title = {Diffuse, sample, project: plug-and-play controllable graph generation},
year = {2024},
booktitle = {ICML},
}

@article{Aur91, 
title={Voronoi Diagrams – A Survey of a Fundamental Geometric Data Structure}, 
volume={23},  
number = {3},
journal={ACM Comput. Surv.}, 
author={F. Aurenhammer}, 
year={1991},  
pages={345–405}
}

@Article{Ort18,
author   = {A. Ortega and P. Frossard and J. Kova\v{c}evi\'{c} and J. Moura and P. Vandergheynst},
title    = {Graph signal processing: Overview, challenges, and applications},
journal  = {Proc. IEEE},
year     = {2018},
volume   = {106},
number   = {5},
pages    = {808-828},
}

@article{JiaFenTay:J23,
year = {2023},
volume = {17},
journal = {Foundations and Trends® in Signal Processing},
title = {Generalizing Graph Signal Processing: High Dimensional Spaces, Models and Structures},
number = {3},
pages = {209-290},
author = {X. Jian and F. Ji and W. P. Tay}
}

@ARTICLE{Sah22,
  author={Sahu, Nitesh and Wu, Linlong and Babu, Prabhu and M. R., Bhavani Shankar and Ottersten, Björn},
  journal={IEEE Trans. Veh. Technol.}, 
  title={Optimal Sensor Placement for Source Localization: A Unified ADMM Approach}, 
  year={2022},
  volume={71},
  number={4},
  pages={4359-4372},
}

@inproceedings{Ma25,
author = {Yuezhou Ma and Haixu Wu and Hang Zhou and Huikun Weng and Jianmin Wang and Mingsheng Long},
title = {{PhySense}: Sensor Placement Optimization for Accurate
Physics Sensing},
year = {2025},
booktitle = {NeurIPS},
}

@article{Ber23,
title = {Adaptive dynamical networks},
journal = {Physics Reports},
volume = {1031},
pages = {1-59},
year = {2023},
author = {Rico Berner and Thilo Gross and Christian Kuehn and Jürgen Kurths and Serhiy Yanchuk},
}

@INPROCEEDINGS{Cha25,
  author={Chandrashekar, Rakesh and Sowjanya, G.Durga and Gupta, Manish and Karuna, Gudapalli and Shareef, Sk Mahammad and Kaushik, Abhishek},
  booktitle={RCSM}, 
  title={Simulation of Autonomous Drone Swarm Coordination and Intelligent Path Planning Leveraging Reinforcement Learning with Geospatial Data Integration}, 
  year={2025},
}

@INPROCEEDINGS{Ale24,
  author={Costa, Alexandre and Duarte, Pedro and Coelho, André and Campos, Rui},
  booktitle={WiMob}, 
  title={Obstacle-Aware Positioning of a Mobile Robotic Platform for Next-Generation Wireless Networks}, 
  year={2024},
}

\end{document}